\documentclass[11pt]{article}
\usepackage{amsmath}
\usepackage{amsthm}
\usepackage{titlesec}
\usepackage{indentfirst}

\theoremstyle{definition}\newtheorem{Df}{Definition}
\theoremstyle{plain}\newtheorem{Th}{Theorem}
\theoremstyle{definition}\newtheorem{Rm}{Remark}
\theoremstyle{definition}
\theoremstyle{plain}
\theoremstyle{plain}\newtheorem{Co}[Th]{Corollary}
\theoremstyle{plain}\newtheorem{Lm}[Th]{Lemma}

\usepackage{amsthm}
\usepackage{amsmath,amssymb}

\begin{document}

\title{{\bf Decidability of the Equivalence of Multi-Letter Quantum Finite Automata}\thanks{This work is supported in
part by the National Natural Science Foundation (Nos.
60873055, 61073054), the
Natural Science Foundation of Guangdong Province of China (No. 10251027501000004), the Fundamental Research Funds for the Central Universities (No. 10lgzd12), the Program
for New Century Excellent Talents in University (NCET) of China,
and the project of  SQIG at IT, funded by FCT and EU FEDER projects
Quantlog
POCI/MAT/55796/2004 and
QSec PTDC/EIA/67661/2006, IT Project QuantTel, NoE Euro-NF, and the SQIG LAP initiative.
}}

\author{Daowen Qiu$^{a,b,c}$ \thanks{Corresponding author (D.W. Qiu): issqdw@mail.sysu.edu.cn} , Xiangfu
Zou$^{a}$, Lvzhou Li$^{a}$, and Paulo Mateus$^{b}$\\
{\small $^{a}$Department of Computer Science, Sun Yat-sen
University, Guangzhou 510006, China}\\
{\small $^{b}$SQIG--Instituto de Telecomunica\c{c}\~{o}es, Departamento de Matem\'{a}tica, Instituto Superior T\'{e}cnico,}\\
{\small  Universidade T\'{e}cnica de Lisboa, Av. Rovisco Pais 1049-001, Lisbon, Portugal}\\
 {\small $^{c}$ The State Key Laboratory of Computer Science, Institute of Software,}\\ {\small Chinese  Academy of Sciences,
 Beijing 100080, China} }

\date{ }

\maketitle

\begin{abstract}
Multi-letter {\it  quantum finite automata} (QFAs) were a quantum variant of classical
{\it one-way multi-head finite automata} (J. Hromkovi\v{c}, Acta Informatica 19 (1983) 377-384), and
it has been shown that this new one-way QFAs (multi-letter QFAs) can accept with no error some
regular languages $(a+b)^{*}b$ that are unacceptable by the previous one-way
QFAs. In this paper, we study the decidability of the equivalence of
multi-letter QFAs, and the main technical contributions are as
follows: (1) We show that any two automata, a $k_{1}$-letter QFA
${\cal A}_1$ and a $k_{2}$-letter QFA ${\cal A}_2$, over the same
input alphabet $\Sigma$ are equivalent if and only if they are
$(n^2m^{k-1}-m^{k-1}+k)$-equivalent, where $m=|\Sigma|$ is the
cardinality of $\Sigma$, $k=\max(k_{1},k_{2})$, and $n=n_{1}+n_{2}$,
with $n_{1}$ and $n_{2}$ being the numbers of states of ${\cal
A}_{1}$ and ${\cal A}_{2}$, respectively.  When $k=1$, we obtain the
decidability of equivalence of measure-once QFAs in the literature. It is worth mentioning that our technical method  is essentially different from that for the decidability of the case of single input alphabet (i.e., $m=1$).
(2) However, if we determine the equivalence of multi-letter QFAs by
checking all strings of length not more than $
n^2m^{k-1}-m^{k-1}+k$, then the worst time complexity is
exponential, i.e., $O(n^6m^{n^2m^{k-1}-m^{k-1}+2k-1})$.  Therefore,
we design a polynomial-time $O(m^{2k-1}n^{8}+km^kn^{6})$ algorithm
for determining the equivalence of any two multi-letter QFAs. Here,
the time complexity is concerning the number of states in the
multi-letter QFAs, and  $k$ is thought of as a constant.


\end{abstract}

\section{Introduction}

Interest in quantum computation and information has steadily increased since
Shor's quantum algorithm for factoring integers in polynomial time
\cite{Shor} and Grover's algorithm of searching in database of
size $n$ with only $O(\sqrt{n})$ accesses \cite{Grover}. As we
know, these algorithms are based on {\it quantum Turing machines}
which seem complicated to implement using today's experiment
technology. Therefore, after it has turned out that making powerful quantum computers is still a long term goal, it gets clear that there is a need to study ``small-size" quantum processors using variations of the models that have shown their relevance in the classical cases \cite{Gru99}.

{\it Quantum finite automata} (QFAs) are a kind of  theoretical
models of quantum computers with finite memory. This kind of
computing machines was first studied  by Moore and Crutchfield
\cite{MC00}, as well as by Kondacs and Watrous \cite{KW97}
independently. Then it was dealt with in depth by Ambainis and
Freivalds \cite{AF98}, Brodsky and Pippenger \cite{BP02}, and other
authors (see, e.g., the references in \cite{Gru99,QL08}). The study
of QFAs is mainly divided into two directions: one is {\it one-way
quantum finite automata} (1QFAs) whose tape heads only move one cell
to the right at each computation step (1QFAs have been extensively
studied  in \cite{BMP03}), and the other is {\it two-way quantum
finite automata} (2QFAs), in which the tape heads are allowed to
move towards the right or left, or to be stationary \cite{KW97}.
(Notably, Amano and Iwama \cite{AI99} dealt with a decidability
problem concerning an intermediate form called 1.5QFAs, whose tape
heads are allowed to move right or to be stationary; Hirvensalo
\cite{Hir07} investigated a decidability problem related to one-way
QFAs.) Furthermore, by considering the number of times the
measurement is performed in a computation, 1QFAs have two different
forms: {\it measure-once} 1QFAs (MO-1QFAs) proposed by Moore and
Crutchfield \cite{MC00}, and, {\it measure-many} 1QFAs (MM-1QFAs)
studied first by Kondacs and Watrous \cite{KW97}.

MM-1QFAs are strictly more powerful than MO-1QFAs
\cite{AF98,BMP03} (Indeed, $a^{*}b^{*}$ can be accepted by
MM-1QFAs with bounded error but not by any MO-1QFA with bounded
error). Due to the unitarity of quantum physics and finite memory
of finite automata, both MO-1QFAs and MM-1QFAs can only accept
proper subclasses of regular languages with bounded error (see,
e.g., \cite{KW97,AF98,BP02,BMP03,Jea07}). Indeed, it was shown that the
regular language $(a+b)^{*}b$ cannot be accepted by any MM-1QFA
with bounded error \cite{KW97}. Here the recognizing fashion of languages is bounded-error, and concerning the unbounded-error cases, Yakaryilmaz and Cem Say \cite{YC08,YC10} have investigated them in detail.

Recently, Belovs, Rosmanis, and Smotrovs \cite{BRS07} proposed a
new one-way QFA model, namely, multi-letter QFAs, which is
thought of as a quantum counterpart of more restricted classical
{\it one-way multi-head finite automata} (see \cite{Hro83} by J. Hromkovi\v{c}, {\it Acta Informatica} 19 (1983) 377-384).
Roughly speaking, a $k$-letter QFA is not limited to seeing only
one, the just-incoming input letter,  but can see several earlier
received letters as well. That is, the quantum state transition
which the automaton performs at each step depends on the last $k$
letters received. For the other computing principle, it is similar
to the MO-1QFAs as described above. Indeed, when $k=1$, it reduces
to an MO-1QFA.  By ${\cal L}(QFA_{k})$ we denote the class of
languages accepted with bounded error by $k$-letter QFAs. Any
given $k$-letter QFA can be simulated by some $(k+1)$-letter QFA.
Qiu and Yu \cite{QY08} have proved  that ${\cal L}(QFA_{k})\subset
{\cal L}(QFA_{k+1})$ for $k=1,2,...$, where the inclusion
$\subset$ is proper. Therefore, $(k+1)$-letter QFAs are
computationally more powerful than $k$-letter QFAs. Belovs et al.
\cite{BRS07} have already showed that $(a+b)^{*}b$ can be accepted
by a 2-letter QFA but, as proved in \cite{KW97}, it cannot be
accepted by any MM-1QFA with bounded error. Therefore,
multi-letter QFAs can accept some regular languages that cannot be
accepted by any MM-1QFA and MO-1QFA.

As we know, determining the equivalence for computing models is an
important issue in the theory of computation (see, e.g.,
\cite{Paz71,Tze92,HHW01}). Two computing models over the same
input alphabet $\Sigma$ are $n$-equivalent if and only if their
accepting probabilities are equal for the input strings of length
not more than $n$, while they are equivalent if and only if their
accepting probabilities are equal for all input strings.

Concerning the problem of determining the equivalence for QFAs,
there exists some work \cite{BP02,MC00} that deals with the
simplest case---MO-1QFAs. For quantum sequential machines (QSMs),
Qiu and Li \cite{Qiu02,LQ06}  gave a method for determining
whether or not any two given QSMs are equivalent. This method
applies to determining the equivalence between any two MO-1QFAs
and also is different from the previous ones. For the equivalence
problem of MM-1QFAs, inspired by the work of \cite {Tze92} and
\cite {BMP03}, Li and Qiu \cite{LQ08} presented a polynomial-time
algorithm for determining whether or not any two given MM-1QFAs
are equivalent. Recently, Qiu and Yu \cite{QY08} proved that any
two automata, a $k_{1}$-letter QFA ${\cal A}_1$ and a
$k_{2}$-letter QFA ${\cal A}_2$ over the same input alphabet
$\Sigma=\{\sigma\}$ are equivalent if and only if they are
$(n_{1}+n_{2})^{4}+k-1$-equivalent, where $n_{1}$ and $n_{2}$ are
the numbers of states of ${\cal A}_{1}$ and ${\cal A}_{2}$,
respectively, and $k=\max(k_{1},k_{2})$. Also, we find that the decidability method in \cite{QY08} is not applied to the decision of equivalence for the case of multi-input alphabet, and a new technical
method is needed to determine the equivalence of multi-letter QFAs for arbitrary
input alphabet $\Sigma$. We will further describe the essential difference of the new method in Remark 2.

Therefore,
in this paper,  we study the equivalence of multi-letter QFAs for finite
input alphabet $\Sigma$ that has arbitrary number of elements, say
$\Sigma=\{\sigma_{1},\sigma_{2},\ldots,\sigma_{m}\}$. More
specifically, we prove that any two automata, a $k_{1}$-letter QFA
${\cal A}_1$ and a $k_{2}$-letter QFA ${\cal A}_2$ over the input
alphabet $\Sigma=\{\sigma_{1},\sigma_{2},\ldots,\sigma_{m}\}$, are
equivalent if and only if they are
$((n_{1}+n_{2})^2m^{k-1}-m^{k-1}+k)$-equivalent, where $n_{1}$ and
$n_{2}$ are the numbers of states of ${\cal A}_{1}$ and ${\cal
A}_{2}$, respectively, $k=\max(k_{1},k_{2})$.  As a corollary, when
$k=1$, we obtain that any two given MO-1QFAs ${\cal A}_1$ and ${\cal
A}_2$ over the input alphabet
$\Sigma=\{\sigma_{1},\sigma_{2},\ldots,\sigma_{m}\}$, are equivalent
if and only if they are $(n_{1}+n_{2})^2$-equivalent. In addition,
when $m=1$, we have that any two automata, a $k_{1}$-letter QFAs
${\cal A}_1$ and a $k_{2}$-letter ${\cal A}_2$, over the input
alphabet $\Sigma=\{\sigma\}$ are equivalent if and only if they are
$((n_{1}+n_{2})^2+k-1)$-equivalent.

However, if we determine the equivalence of multi-letter QFAs by
checking all strings $x$ with $|x|\leq n^2m^{k-1}-m^{k-1}+k$, then
the time complexity is exponential, i.e.,
$O(n^{6}m^{n^2m^{k-1}-m^{k-1}+2k-1})$, where $|x|$ is the length of
$x$, and $n=n_{1}+n_{2}$. Therefore, in this paper, we design a
polynomial-time $O(m^{2k-1}n^{8}+km^kn^{6})$ algorithm for
determining the equivalence of any two multi-letter QFAs.

The remainder of the paper is organized as follows. In Section 2, we
recall the definition of multi-letter QFAs and other related
definitions, and some related results are reviewed. In Section 3, we
give a condition (mentioned before) for whether two multi-letter
QFAs are equivalent. Also, some corollaries are obtained, and those
corollaries correspond to results found in literature. Then, in
Section 4, we design a polynomial-time $O(m^{2k-1}n^{8}+km^kn^{6})$
algorithm for determining the equivalence of any two multi-letter
QFAs. Finally, in Section 5 we address some related issues for
further consideration.

We would mention that the technical process in this paper is new
and much more complicated than that in the literature
\cite{MC00,BP02,LQ06,LQ08,LQ08N} regarding the equivalence of
1QFAs. In general, symbols will be explained when they first
appear. In this paper, for a matrix $A$, we use $A^*$ and
$A^\dagger$ to denote its conjugate and conjugate transpose,
respectively.

\section{Preliminaries}

In this section, we briefly review some definitions and related
properties that will be used in the consequent sections. For the
details, we refer to \cite{BRS07,QY08}. First we recall $k$-letter
deterministic finite automata ($k$-letter DFAs).

\begin{Df}[{\cite{BRS07}}]
 A $k$-letter {\it deterministic finite automaton} ($k$-letter DFA) is defined by a
 quintuple $(Q,Q_{acc},q_{0},\Sigma,\gamma)$, where $Q$ is a
 finite set of states, $Q_{acc}\subseteq Q$ is the set of
 accepting states, $q_{0}\in Q$ is the initial state, $\Sigma$ is
 a finite input alphabet, and $\gamma$ is a transition
 function that maps $Q\times T^{k}$ to $Q$, where
 $T=\{\Lambda\} \bigcup \Sigma$ and letter $\Lambda\notin \Sigma$
 denotes the empty letter, and $T^{k}\subset T^{*}$ consists of all strings of length
 $k$.
\end{Df}

We describe the computing process of a $k$-letter DFA on an input
string $x$ in $\Sigma^{*}$, where
$x=\sigma_{1}\sigma_{2}\cdots\sigma_{n}$, and $\Sigma^{*}$ denotes
the set of all strings over $\Sigma$.  The $k$-letter DFA has a
tape which contains the letter $\Lambda$ in its first $k-1$
positions followed by the input string $x$. The automaton starts
in the initial state $q_{0}$ and has $k$ reading heads which
initially are on the first $k$ positions of the tape (clearly, the
$k$th head reads $\sigma_{1}$ and the other heads read $\Lambda$).
Then the automaton transfers to a new state as the current state and
all heads move right a position in parallel. Now the $(k-1)$th and
$k$th heads point to $\sigma_{1}$ and $\sigma_{2}$, respectively,
and the others, if any, to $\Lambda$. Subsequently, the automaton
transfers to a new state and all heads move to the right. This
process does not stop until the $k$th head has read the last
letter $\sigma_{n}$. The input string $x$ is accepted if and only
if the automaton enters an accepting state after its $k$th head
reading the last letter $\sigma_{n}$.

Clearly, $k$-letter DFAs are not more powerful than DFAs. Indeed,
the family of languages accepted by $k$-letter DFAs, for $k\geq
1$, is exactly the family of regular languages. A {\it group
finite automaton} (GFA) \cite{BP02} is a DFA whose state
transition function, say $\delta$, satisfies that for any input
symbol $\sigma$, $\delta(\cdot,\sigma)$ is a one-to-one map on the
state set, i.e., a permutation on the state set.

\begin{Df}[{\cite{BRS07}}]
 A $k$-letter DFA $(Q,Q_{acc},q_{0},\Sigma,\gamma)$ is called a $k$-letter group finite automaton
 ($k$-letter GFA) if and only if for any string $x\in T^{k}$ the
 function $\gamma_{x}(q)=\gamma(q,x)$ is a bijection from
 $Q$ to $Q$.
\end{Df}

Now we recall the definition of multi-letter QFAs \cite{BRS07}.

\begin{Df} [{\cite{BRS07}}]
A $k$-letter QFA ${\cal A}$ is defined by a quintuple ${\cal
A}=(Q,Q_{acc},|\psi_{0}\rangle, \Sigma,\mu)$ where $Q$ is a set of
states, $Q_{acc}\subseteq Q$ is the set of accepting states,
$|\psi_{0}\rangle$ is the initial unit state that is a
superposition of the states in $Q$, $\Sigma$ is a finite input
alphabet, and $\mu$ is a function that assigns a unitary
transition matrix $U_{w}$ on $\mathbb{C}^{|Q|}$ for each string
$w\in (\{\Lambda\}\cup\Sigma)^{k}$, where $|Q|$ is the cardinality
of $Q$.

\end{Df}

The computation of a $k$-letter QFA ${\cal A}$ works in the same
way as the computation of an MO-1QFA \cite{MC00,BP02}, except that
it applies unitary transformations corresponding not only to the
last letter but the last $k$ letters received (like a $k$-letter
DFA). When $k=1$, it is exactly an MO-1QFA as pointed out before.
According to \cite{BRS07}, all languages accepted by $k$-letter
QFAs with bounded error are regular languages for any $k$.

Now we give the probability $P_{{\cal A}}(x)$ for $k$-letter QFA
${\cal A}=(Q,Q_{acc},|\psi_{0}\rangle, \Sigma,\mu)$ accepting any
input string $x=\sigma_{1}\sigma_{2}\cdots\sigma_{m}$. From the
definition we know that, for any $w\in
(\{\Lambda\}\cup\Sigma)^{k}$, $\mu(w)$ is a unitary matrix. By the
definition of $\mu$, we can define the unitary transition for each
string $x=\sigma_{1}\sigma_{2}\cdots\sigma_{m}\in\Sigma^{*}$. By
$\overline{\mu}$ (induced by $\mu$) we mean a map from
$\Sigma^{*}$ to the set of all $|Q|\times |Q|$ unitary matrices.
Specifically, $\overline{\mu}$ is induced by $\mu$ in the
following way: For
$x=\sigma_{1}\sigma_{2}\cdots\sigma_{m}\in\Sigma^{*}$,
\begin{equation}
\overline{\mu}(x)=\left\{\begin{array}{ll}
\mu(\Lambda^{k-1}\sigma_{1})\mu(\Lambda^{k-2}\sigma_{1}\sigma_{2})\cdots\mu(\Lambda^{k-m}x), &  {\rm if} \  m<k, \\
\mu(\Lambda^{k-1}\sigma_{1})\mu(\Lambda^{k-2}\sigma_{1}\sigma_{2})\cdots\mu(\sigma_{m-k+1}\sigma_{m-k+2}\cdots\sigma_{m}),
& {\rm if}\  m\geq k,
   \end{array}
 \right.
\end{equation}
which implies the computing process of ${\cal A}$ for input string
$x$.

We identify the states in $Q$ with an orthonormal basis of
$\mathbb{C}^{|Q|}$, and let $P_{acc}$ denote the projection
operator on the subspace spanned by $Q_{acc}$. Then we define that
\begin{equation}
P_{{\cal A}}(x)=\|\langle\psi_{0}|\overline{\mu}(x)P_{acc}\|^{2}.
\end{equation}

\section{ Determining the equivalence between multi-letter
quantum finite automata}

For any given $k_{1}$-letter QFA ${\cal A}_{1}$ and $k_{2}$-letter
QFA ${\cal A}_{2}$ over the same input alphabet
$\Sigma=\{\sigma_{1},\sigma_{2},\ldots,\sigma_{m}\}$, our purpose
is to determine whether they are equivalent. We give the
definition of equivalence between two multi-letter QFAs as follows.

\begin{Df}
A $k_{1}$-letter QFA ${\cal A}_1$ and another $k_{2}$-letter QFA
${\cal A}_2$  over the same input alphabet $\Sigma$ are said to be
equivalent (resp. $t$-equivalent) if $P_{{\cal A}_1}(w)=P_{{\cal
A}_2}(w)$ for any $w\in \Sigma^{*} $ (resp. for any input string
$w$ with $ |w|\leq t$).
\end{Df}

We introduce some notation. $\mathbb{C}^{n}$ denotes the
Euclidean space consisting of all $n$-dimensional complex vectors.
 For a subset $S$ of
$\mathbb{C}^{n}$, $\textrm{span} S$ represents the minimal subspace spanned
by $S$. For any given $k$-letter QFA ${\cal
A}=(Q,Q_{acc},|\psi_{0}\rangle, \Sigma,\mu)$, we denote $
\mathbb{F}(j) =\textrm{span}
\{\langle\psi_0|\overline{\mu}(x):x\in\Sigma^*, |x|\leq j\}$,
$j=1,2,\cdots$. That is, $\mathbb{F}(j)$ is a subspace of
$\mathbb{C}^{|Q|}$, spanned by
$\{\langle\psi_0|\overline{\mu}(x):x\in\Sigma^*, |x|\leq j\}$,
where $|Q|$ denotes the  the number of states of $Q$.

\begin{Lm}\label{Th1} Let  ${\cal A}=(Q,Q_{acc},|\psi_{0}\rangle, \Sigma,\mu)$ be a
$k$-letter QFA, where $\Sigma=\{\sigma_i:i=1,2,\cdots,m\}$. Then
there exists an integer $i_0\leq  (n-1)m^{k-1}+k$ such that, for
any $i\geq i_0$, $ \mathbb{F}(i) = \mathbb{F}(i_0) $, where $n$ is
the number of states of $Q$.
\end{Lm}

\noindent\textbf{Proof.} We denote
\begin{equation}
\Sigma^{(k-1)}=\{x:x\in\Sigma^*,|x|=k-1\},
\end{equation}
and
\begin{equation}
\mathbb{G}(l,w)=\textrm{span}\{\langle\psi_0|\overline{\mu}(xw)
:x\in\Sigma^*,|x|\leq l\}
\end{equation}
for any $ w\in \Sigma^{(k-1)}$ and any $l\in\{0,1,2,\cdots\}$. In
addition,  we denote
\begin{equation}
\mathbb{H}(l)=\oplus_{w\in \Sigma^{(k-1)}} \mathbb{G}(l,w)
\end{equation}
for any $l\in\{0,1,2,\cdots\}$, where $\oplus_{w\in
\Sigma^{(k-1)}} \mathbb{G}(l,w)$ is the direct sum of
$\mathbb{G}(l,w)$ for all $w\in \Sigma^{(k-1)}$. It is clear that
\begin{equation}
\mathbb{G}(l,w)\subseteq \mathbb{G}(l+1,w),\;\forall w\in
\Sigma^{(k-1)} \;\text{and} \;\forall
l\in\{0,1,2,\cdots\},\end{equation} and
\begin{equation}
\mathbb{H}(l)\subseteq  \mathbb{H}(l+1),\;\forall
l\in\{0,1,2,\cdots\}. \label{H}
\end{equation}
Since $\mathbb{G}(l,w)$ is a subspace of $\mathbb{C}^{n}$, we
obtain that
\begin{equation}
1\leq \dim(\mathbb{G}(l,w))\leq n,\end{equation}  for any $ w\in
\Sigma^{(k-1)}$ and any $l\in\{0,1,2,\cdots\}$, where
$\dim(\mathbb{G}(l,w))$ denotes the dimension of
$\mathbb{G}(l,w)$. Furthermore, by the definition of direct sum,
we have
\begin{equation}
 m^{k-1} \leq \dim(\mathbb{H}(l))\leq n m^{k-1}\end{equation}
for  any $l\in\{0,1,2,\cdots\}$. Therefore, by Eq. (\ref{H}) there
exists $l_0\leq (n-1)m^{k-1}+1$ such that
\begin{equation}
\mathbb{H}(l_0)=\mathbb{H}(l_0+1).\end{equation} Equivalently,
\begin{equation}
\mathbb{G}(l_0,w)=\mathbb{G}(l_0+1,w),\;\,\forall w\in
\Sigma^{(k-1)}.\end{equation}

 Let $i_0=l_0+(k-1)\leq  (n-1)m^{k-1}+k$. Now, we prove by
induction that, for any $i\geq i_0$, $ \mathbb{F}(i) =
\mathbb{F}(i_0) $.

\emph{Base step}. When $i=i_0$, it is clear that $ \mathbb{F}(i) =
\mathbb{F}(i_0) $.

\emph{Induction step}. Suppose $ \mathbb{F}(j) = \mathbb{F}(i_0) $,
for some $j\geq i_0$. Our purpose is to prove that $ \mathbb{F}(j+1)
= \mathbb{F}(i_0) $.  For any given $w\in\Sigma^{(j+1)}$, we denote
$w=\sigma_1\sigma_2\cdots\sigma_{l_0+1}\sigma_{l_0+2}\cdots
\sigma_{i_0}\sigma_{i_0+1}\cdots\sigma_j\sigma_{j+1}$ and let
$w_0=\sigma_{l_0+2}\cdots \sigma_{i_0}\sigma_{i_0+1}$. Clearly,
$w_{0}\in \Sigma^{(k-1)}$, and
$\langle\psi_0|\overline{\mu}(\sigma_1\sigma_2\cdots\sigma_{l_0+1}\sigma_{l_0+2}\cdots
\sigma_{i_0}\linebreak[0]\sigma_{i_0+1})\linebreak[0]=
\langle\psi_0|\overline{\mu}(\sigma_1\sigma_2\cdots\sigma_{l_0+1}
w_{0})\in \mathbb{G}(l_0+1,w_{0})$.

Due to $\mathbb{H}(l_0)=\mathbb{H}(l_0+1)$, i.e.,
$\mathbb{G}(l_0,w)=\mathbb{G}(l_0+1,w)$ for any $ w\in
\Sigma^{(k-1)}$, we obtain that
$\langle\psi_0|\overline{\mu}(\sigma_1\sigma_2\cdots\sigma_{l_0+1}
w_{0})\in \mathbb{G}(l_0,w_{0})$. Therefore,
$\langle\psi_0|\overline{\mu}(\sigma_1\sigma_2\cdots\sigma_{l_0+1}
w_{0})$ can be linearly represented by the vectors of
$\mathbb{G}(l_0,w_{0})$. As a result, there exist a finite index
set $\Gamma$ and $x_{\gamma}\in \{x:x\in\Sigma^*,\;|x|\leq l_0\}$
as well as complex numbers $p_\gamma$ with $\gamma\in\Gamma$, such
that
\begin{equation}
\langle\psi_0|\overline{\mu}(\sigma_1\sigma_2\cdots\sigma_{l_0+1}w_{0})
=\sum_{\gamma\in\Gamma} p_\gamma
\langle\psi_0|\overline\mu(x_\gamma w_0) .\end{equation}
Therefore, we have
\begin{eqnarray*}
\langle\psi_0|\overline{\mu}(w)&=&\langle\psi_0|\overline{\mu}(\sigma_1\sigma_2\cdots\sigma_{l_0+1}
w_{0}\sigma_{i_{0}+2}\cdots\sigma_{j+1})\\
&=&\langle\psi_0|\overline{\mu}(\sigma_1\sigma_2\cdots\sigma_{l_0+1}
w_{0}) \mu(w_{0}\sigma_{i_{0}+2})\cdots\mu(\sigma_{j-k+2}\cdots\sigma_{j+1})\\
&=&\sum_{\gamma\in\Gamma} p_{\gamma}
\langle\psi_0|\overline\mu(x_\gamma w_0)\mu(w_{0}\sigma_{i_{0}+2})\cdots\mu(\sigma_{j-k+2}\cdots\sigma_{j+1})\\
&=& \sum_{\gamma\in\Gamma} p_\gamma
\langle\psi_0|\overline\mu(x_\gamma \sigma_{l_0+2}\cdots
\sigma_{i_0}\sigma_{i_0+1}\cdots\sigma_j\sigma_{j+1})\in
\mathbb{F}(j).
\end{eqnarray*}
Consequently, $\langle\psi_0|\overline{\mu}(w)\in \mathbb{F}(j)$,
and we get that $ \mathbb{F}(j+1) \subseteq \mathbb{F}(j)$. On the
other hand,  $ \mathbb{F}(j)\subseteq \mathbb{F}(j+1) $ always
holds. Hence, we obtain that $ \mathbb{F}(j+1) =
\mathbb{F}(j)=\mathbb{F}(i_{0})$. The proof is completed. \qed\\

With the same method of proof as that for Lemma 1, we can obtain
the following lemma.

\begin{Lm}\label{Th2}
For $\Sigma=\{\sigma_i:i=1,2,\cdots,m\}$ and a $k$-letter QFA
${\cal A}=(Q,Q_{acc}, |\psi_0\rangle,\Sigma,\mu)$, there exists an
integer $i_0\leq  (n^2-1)m^{k-1}+k$ such that, for any $i\geq
i_0$, $ \mathbb{E}(i) = \mathbb{E}(i_0) $, where $n$ is the number
of states of $Q$, $ \mathbb{E}(j)
=\textrm{span}\{\langle\psi_{0}|\otimes(\langle\psi_{0}|)^{*}
\overline\nu(x): x\in \Sigma^{*}, |x|\leq j\}$ for $j=1,2,\cdots$,
and $\overline\nu(x)=
\overline{\mu}(x)\otimes\overline{\mu}(x)^{*}$, where $*$ denotes
the conjugate operation.
\end{Lm}

\noindent\textbf{Proof.}  It is exactly similar to the proof of
Lemma \ref{Th1}. \qed\\

To prove the equivalence of multi-letter QFAs, we further present
a lemma. For $\Sigma=\{\sigma_1,\,\sigma_2,\,\cdots,\,\sigma_{m}
\}$, a $k_{1}$-letter QFA ${\cal
A}_1=(Q_{1},Q_{acc}^{(1)},|\psi_{0}^{(1)}\rangle, \Sigma,\mu_{1})$
and another $k_{2}$-letter QFA ${\cal
A}_2=(Q_{2},Q_{acc}^{(2)},|\psi_{0}^{(2)}\rangle,
\Sigma,\mu_{2})$, let $P_{acc}^{(1)}$ and $P_{acc}^{(2)}$ denote
the projection operators on the subspaces spanned by
$Q_{acc}^{(1)}$ and $Q_{acc}^{(2)}$, respectively. For any string
$x\in\Sigma^{*}$, we set
$\overline{\mu}(x)=\overline{\mu}_{1}(x)\oplus\overline{\mu}_{2}(x)$
and $P_{acc}=P_{acc}^{(1)}\oplus P_{acc}^{(2)}$,
$Q_{acc}=\{|\eta_1\rangle\oplus|\eta_2\rangle:|\eta_1\rangle\in
Q_{acc}^{(1)} \text{ and } |\eta_2\rangle\in Q_{acc}^{(2)}\}$. In
addition, we denote $|\eta_{1}\rangle=|\psi_{0}^{(1)}\rangle\oplus
{\bf 0_{2}}$ and $|\eta_{2}\rangle={\bf 0_{1}}\oplus
|\psi_{0}^{(2)}\rangle$, where ${\bf 0_{1}}$ and ${\bf 0_{2}}$
represent column zero vectors of $n_{1}$ and $n_{2}$ dimensions,
respectively.

\begin{Lm}\label{Th3}
${\cal A}_1$ and ${\cal A}_2$ above are equivalent if and only if
\begin{eqnarray}\label{Eq2}
(\langle\eta_{1}|(\langle\eta_{1}|)^{*}
-\langle\eta_{2}|(\langle\eta_{2}|)^{*})
\overline\nu(x)\sum_{p_{j}\in
Q_{acc}}(|p_{j}\rangle(|p_{j}\rangle)^{*})=0 \label{EqLm3}
\end{eqnarray}
for all strings $x\in\Sigma^{+}$, where $\overline\nu(x)=\overline
\mu(x)\otimes\overline\mu(x)^*$.
\end{Lm}

\noindent\textbf{Proof.} Denote that, for any string $x\in
\Sigma^{*}$,
\begin{equation}
P_{\eta_{1}}(x)=\|\langle\eta_{1}|\overline{\mu}(x)P_{acc}\|^{2}
\end{equation}
and
\begin{equation}
P_{\eta_{2}}(x)=\|\langle\eta_{2}| \overline{\mu}
(x)P_{acc}\|^{2}.
\end{equation}
Indeed, we further have that
\begin{eqnarray}
\nonumber P_{\eta_{1}}(x)&=&\|\langle\eta_{1}|\overline{\mu}(x)P_{acc}\|^{2}\\
&=&\nonumber\langle\eta_{1}|\overline{\mu}(x)P_{acc}P_{acc}^{\dagger}\overline{\mu}(x)^{\dagger}|\eta_{1}\rangle\\
&=&\nonumber\langle\eta_{1}|\overline{\mu}(x)P_{acc}\overline{\mu}(x)^{\dagger}|\eta_{1}\rangle\\
&=&\nonumber\langle\psi_{0}^{(1)}|\overline{\mu}_{1}(x)P_{acc}^{(1)}\overline{\mu}_{1}(x)^{\dagger}|\psi_{0}^{(1)}\rangle\\
&=&P_{{\cal A}_{1}}(x).
\end{eqnarray}
Similarly,
\begin{equation}
P_{\eta_{2}}(x)=P_{{\cal A}_{2}}(x).
\end{equation}
Therefore,
\begin{equation}\label{Eq0}
P_{{\cal A}_{1}}(x)=P_{{\cal A}_{2}}(x)
\end{equation}
holds if and only if
\begin{equation}\label{Eq1}
P_{\eta_{1}}(x)=P_{\eta_{2}}(x)
\end{equation}
for any string $x\in \Sigma^{*}$. On the other hand, we have that
\begin{eqnarray}
\nonumber P_{\eta_{1}}(x)&=&\|\langle\eta_{1}|\overline{\mu}(x)P_{acc}\|^{2}\\
&=&\nonumber\sum_{p_{j}\in Q_{acc}}|\langle\eta_{1}|\overline{\mu}(x)|p_{j}\rangle|^{2}\\
&=&\nonumber\sum_{p_{j}\in Q_{acc}}\langle\eta_{1}|\overline{\mu}(x)|p_{j}\rangle(\langle\eta_{1}|\overline{\mu}(x)|p_{j}\rangle)^{*}\\
&=&\nonumber\sum_{p_{j}\in Q_{acc}}\langle\eta_{1}|(\langle\eta_{1}|)^{*}(\overline{\mu}(x)\otimes (\overline{\mu}(x))^{*})|p_{j}\rangle(|p_{j}\rangle)^{*}\\
&=&\langle\eta_{1}|(\langle\eta_{1}|)^{*}(\overline{\mu}(x)\otimes
(\overline{\mu}(x))^{*})\sum_{p_{j}\in
Q_{acc}}|p_{j}\rangle(|p_{j}\rangle)^{*}.
\end{eqnarray}
Similarly,
\begin{eqnarray}
P_{\eta_{2}}(x)=\langle\eta_{2}|(\langle\eta_{2}|)^{*}(\overline{\mu}(x)\otimes
(\overline{\mu}(x))^{*})\sum_{p_{j}\in
Q_{acc}}|p_{j}\rangle(|p_{j}\rangle)^{*}.
\end{eqnarray}
Therefore, Eq. (\ref{Eq1}) holds if and only if
\begin{eqnarray}
&&\nonumber\langle\eta_{1}|(\langle\eta_{1}|)^{*}(\overline{\mu}(x)\otimes
(\overline{\mu}(x))^{*})\sum_{p_{j}\in
Q_{acc}}|p_{j}\rangle(|p_{j}\rangle)^{*}\\
&=&\langle\eta_{2}|(\langle\eta_{2}|)^{*}(\overline{\mu}(x)\otimes
(\overline{\mu}(x))^{*})\sum_{p_{j}\in
Q_{acc}}|p_{j}\rangle(|p_{j}\rangle)^{*} \label{Eq}
\end{eqnarray}
for any string $x\in \Sigma^{*}$. Denote
\begin{equation}
\overline\nu(x)= \overline{\mu}(x)\otimes\overline{\mu}(x)^{*}.
\end{equation}
Clearly $\overline\nu(x)$ is  an $n^{2}\times n^{2}$ complex
square matrix. Then the equivalence between ${\cal A}_1$ and
${\cal A}_2$ depends on whether or not the following equation
holds for any string $x\in\Sigma^{*}$:
\begin{eqnarray}
\langle\eta_{1}|(\langle\eta_{1}|)^{*}
\overline\nu(x)\sum_{p_{j}\in
Q_{acc}}|p_{j}\rangle(|p_{j}\rangle)^{*}=\langle\eta_{2}|(\langle\eta_{2}|)^{*}
\overline\nu(x)\sum_{p_{j}\in
Q_{acc}}|p_{j}\rangle(|p_{j}\rangle)^{*},
\end{eqnarray}
i.e.,
\begin{eqnarray}\label{Eq2}
(\langle\eta_{1}|(\langle\eta_{1}|)^{*}
-\langle\eta_{2}|(\langle\eta_{2}|)^{*}) \overline\nu(x)
\sum_{p_{j}\in Q_{acc}}|p_{j}\rangle(|p_{j}\rangle)^{*}=0.
\end{eqnarray}
\qed\\

With the above lemmas we are ready to prove the main theorem.

\begin{Th}\label{Th3}
${\cal A}_1$ and ${\cal A}_2$ above are equivalent if and only if
they are $( n^2m^{k-1}-m^{k-1}+k)$-equivalent, where
$k=\max(k_{1},k_{2})$ and $n=n_1+n_2$, with $n_{i}$ being the
number of states in $Q_{i}$, $i=1,2$.
\end{Th}

\noindent\textbf{Proof.} Denote $\mathbb{D}(j)
=span\{(\langle\eta_{1}|(\langle\eta_{1}|)^{*}
-\langle\eta_{2}|(\langle\eta_{2}|)^{*}) \overline\nu(x):
x\in\Sigma^{*}, |x|\leq j\}$ for $j=1,2,\cdots$. Here
$\mathbb{D}(j)$ is a subspace of $\mathbb{C}^{n^{2}}$. By Lemma
\ref{Th2}, we can readily obtain that there exists an $i_0\leq
(n^2-1)m^{k-1}+k$ such that
\begin{equation}\label{Eq3}
\mathbb{D}(i) = \mathbb{D}(i_0)
\end{equation}
for all $i\geq i_0$.  Eq. (\ref{Eq3}) implies that, for any
$x\in\Sigma^{*}$ with $|x|> (n^2-1)m^{k-1}+k$,
$(\langle\eta_{1}|(\langle\eta_{1}|)^{*}
-\langle\eta_{2}|(\langle\eta_{2}|)^{*}) \overline\nu(x)$ can be
linearly represented by some vectors in
$\{(\langle\eta_{1}|(\langle\eta_{1}|)^{*}
-\langle\eta_{2}|(\langle\eta_{2}|)^{*}) \overline\nu(y):
y\in\Sigma^{*}\; \text{and}\; |y|\leq (n^2-1)m^{k-1}+k\}$.

Consequently, if Eq. (\ref{EqLm3}) holds for all $x$ with $|x|\leq
(n^2-1)m^{k-1}+k$, then so does it for all $x$ with
$|x|>(n^2-1)m^{k-1}+k$. We have proved this theorem. \qed\\

From Theorem \ref{Th3} we can get a sufficient and necessary
condition for the equivalence of MO-1QFAs ($k=1$), and we describe
it by the following corollary, which was also presented by Li and
Qiu \cite{LQ08N}.

\begin{Co}
For $\Sigma=\{\sigma_1,\,\sigma_2,\,\cdots,\,\sigma_{m} \}$, an
MO-1QFA ${\cal A}_1=(Q_{1},Q_{acc}^{(1)},|\psi_{0}^{(1)}\rangle,
\Sigma,\mu_{1})$ and another MO-1QFA ${\cal
A}_2=(Q_{2},Q_{acc}^{(2)},|\psi_{0}^{(2)}\rangle, \Sigma,\mu_{2})$
are equivalent if and only if they are $(n_1+n_2)^2$-equivalent,
where $n_{i}$ is the number of states of $Q_{i}$, $i=1,2$.
\end{Co}

\begin{Rm}
We analyze the complexity of computation in Theorem \ref{Th3}. As
in \cite{Tze92}, we assume that all the inputs consist of complex
numbers whose real and imaginary parts are rational numbers and
that each arithmetic operation on rational numbers can be done in
constant time. Again we denote $n=n_{1}+n_{2}$.  Note that in time
$O(in^{4})$ we check whether or not Eq. (\ref{EqLm3}) holds for
$x\in \Sigma^{*}$ with $|x|=i$. Because the length of $x$ to be
checked in Eq. (\ref{EqLm3}) is at most $(n^2-1)m^{k-1}+k$, the
time complexity for checking whether the two multi-letter QFAs are
equivalent is $O(n^4 (m + 2m^2 + ... +
((n^2-1)m^{k-1}+k)m^{n^2m^{k-1}-m^{k-1}+k}))$, that is
$O(n^6m^{n^2m^{k-1}-m^{k-1}+2k-1})$.

\end{Rm}

From Theorem \ref{Th3} we can obtain a sufficient and necessary
condition for the equivalence of multi-QFAs over the same single
input alphabet, and we describe it by the following corollary,
which has been studied by Qiu and Yu \cite{QY08}.

\begin{Co}
For $\Sigma=\{\sigma\}$, a $k_{1}$-letter QFA ${\cal
A}_1=(Q_{1},Q_{acc}^{(1)},|\psi_{0}^{(1)}\rangle, \Sigma,\mu_{1})$
and another $k_{2}$-letter QFA ${\cal
A}_2=(Q_{2},Q_{acc}^{(2)},|\psi_{0}^{(2)}\rangle, \Sigma,\mu_{2})$
are equivalent if and only if they are $(n^2+k-1)$-equivalent,
where $k=\max(k_{1},k_{2})$ and $n=n_1+n_2$, with $n_{i}$ being
the number of states of $Q_{i}$, $i=1,2$.
\end{Co}

\begin{Rm}

An essential difference for determining the equivalence of
$k$-letter QFAs between ``multi-input alphabet'' and ``single input alphabet'' is the following
fact.

When we study a $k$-letter QFA ${\cal
A}=(Q,Q_{acc},|\psi_{0}\rangle, \Sigma,\mu)$ with multi-input alphabet (i.e., $|\Sigma|\geq 2$), we can not always
obtain
\begin{equation}
\langle\psi_0|\overline{\mu}(x\sigma)=\sum_{\gamma\in\Gamma}
c_\gamma\langle\psi_0|\overline{\mu}(y_\gamma\sigma)\label{EQU1}
\end{equation}
from
\begin{equation}
\langle\psi_0|\overline{\mu}(x)=\sum_{\gamma\in\Gamma}
c_\gamma\langle\psi_0|\overline{\mu}(y_\gamma)\label{EQU2}
\end{equation}
where $x, y_\gamma\in \Sigma^*$, $\sigma\in \Sigma$, and $c_\gamma$
is a complex number.
 Furthermore, we can not obtain
\begin{equation}
\langle\psi_0|\overline{\mu}(x z)=\sum_{\gamma\in\Gamma}
c_\gamma\langle\psi_0|\overline{\mu}(y_\gamma z)
\end{equation}
from
Eq.~(\ref{EQU2})
where $z\in \Sigma^*$.

However, we can obtain Eq.~(\ref{EQU1}) from Eq.~(\ref{EQU2}) for single input alphabet.
In general, if Eq.~(\ref{EQU1}) can be obtained from
Eq.~(\ref{EQU2}), we must let the strings of the last $k-1$ letters
of $y_\gamma, \gamma\in\Gamma,$ be the same as those of $x$.
Therefore, we divided $\{x: x\in\Sigma^* \text{ and }|x|\geq k-1\}$
into $m^{k-1}$ classes by the last $k-1$ letters in our paper.

\end{Rm}

\section{ A polynomial-time algorithm for determining the equivalence between multi-letter
QFAs}

According to our analysis of the complexity of computation in
Remark 1, we need exponential time for checking whether or not two
multi-letter QFAs are equivalent if Eq. (\ref{EqLm3}) is checked
for all strings $x\in \Sigma^{*}$ with $|x|\leq (n^2-1)m^{k-1}+k$.
In this section, our purpose is to design a polynomial-time
algorithm for determining the equivalence between any two
multi-letter QFAs.

We still use the symbols from Lemma 3 and Theorem \ref{Th3} and
its proof. In addition, denote
$\langle\eta|=\langle\eta_{1}|(\langle\eta_{1}|)^{*}-\langle\eta_{2}|(\langle\eta_{2}|)^{*}$
and $|P_{acc}\rangle=\sum_{p_{j}\in
Q_{acc}}|p_{j}\rangle(|p_{j}\rangle)^{*}$. Then, Eq. (\ref{EqLm3})
is equivalent to
\begin{equation}\label{P1}
\langle\eta|\overline\nu(x)|P_{acc}\rangle=0.
\end{equation}
For $w\in\Sigma^{(k-1)}$, denote $\mathbb{G}(l, w)
=span\{\langle\eta| \overline\nu(xw): x\in\Sigma^{*}, |x|\leq
l\}$. Then, from the proof of Lemma 1 we know that there exists
$l_{0}\leq (n^{2}-1)m^{k-1}+1$ such that, for any
$w\in\Sigma^{(k-1)}$ and any $l\geq l_{0}$, $\mathbb{G}(l,
w)=\mathbb{G}(l_{0}, w)$. Since $\mathbb{G}(l_{0}, w)$ is a
subspace of $\mathbb{C}^{n^{2}}$, $\dim(\mathbb{G}(l_{0}, w))\leq
n^{2}$.

If within polynomial time we can find a basis, denoted by ${\cal
B}(w)$, of the subspace $\mathbb{G}(l_{0}, w)$, for all
$w\in\Sigma^{(k-1)}$, then any element in
$\cup_{w\in\Sigma^{(k-1)}}\mathbb{G}(l_{0}, w)$ can be linearly
represented by these elements in $\cup_{w\in\Sigma^{(k-1)}}{\cal
B}(w)$. Therefore, to determine  whether or not  Eq. (\ref{EqLm3})
holds, it suffices to check Eq. (\ref{EqLm3}) for all elements in $
\cup_{w\in\Sigma^{(k-1)}}{\cal B}(w)\bigcup
\{\langle\eta|\overline\nu(x): x\in \cup_{l=0}^{k-2}\Sigma^{(l)}\}$.

\subsection{A polynomial-time algorithm for finding out a base of $\mathbb{G}(l_{0}, w)$}

First, we give some useful definitions. For
$\Sigma=\{\sigma_1,\,\sigma_2,\,\cdots,\,\sigma_m \}$, we define a
\emph{strict order}\cite{R08} $``<"$ on $\Sigma^+$ ($\Sigma^+
=\Sigma^* \setminus \{\epsilon$\}) by, $\forall x_1,x_2\in
\Sigma^+$,
\begin{equation}
x_1<x_2 \text{ iff } |x_1|<|x_2| \text{ or } (|x_1|=|x_2|,
x_1=y\sigma_i z_1, x_2=y\sigma_j z_2 \text{ and } i<j),
\end{equation}
where $\sigma_i,\sigma_j\in\Sigma$ and $y,z_1,z_2 \in \Sigma^*$.
Based on the strict order $ < $, we define a \emph{partial
order}\cite{R08} $``\leq"$ on $\Sigma^+$ by, $\forall x_1,x_2\in
\Sigma^+$,
\begin{equation}
x_1\leq x_2 \text{ iff } x_1<x_2 \text{ or } x_1=x_2.
\end{equation}
The partial order $\leq$ can be expanded to $\Sigma^*$ if we set
$\epsilon\leq x$ for all $x\in\Sigma^*$. Clearly, $\Sigma^*$ is a
well-ordered set \cite{R08}, i.e., every nonempty subset of
$\Sigma^*$ contains a least element.

In the following we design a polynomial-time algorithm to search
for ${\cal B}(w)$ for all $w\in\Sigma^{(k-1)}$, which is described
in the following theorem.

\begin{Th}

For a $k_{1}$-letter QFA ${\cal
A}_1=(Q_{1},Q_{acc}^{(1)},|\psi_{0}^{(1)}\rangle, \Sigma,\mu_{1})$
and another $k_{2}$-letter QFA ${\cal
A}_2=(Q_{2},Q_{acc}^{(2)},|\psi_{0}^{(2)}\rangle, \Sigma,\mu_{2})$,
where $\Sigma=\{\sigma_1,\,\sigma_2,\,\cdots,\,\sigma_{m} \}$, there
exists a polynomial time $O(m^{2k-1}n^{8}+km^kn^{6})$ algorithm
searching for $\bigcup_{w\in\Sigma^{(k-1)}}{\cal B}(w)$, where, as above,
${\cal B}(w)$ is a base of subspace $\mathbb{G}(l_{0}, w)$,
 $l_{0}\leq (n^{2}-1)m^{k-1}+1$, and $n=n_1+n_2$, with $n_{i}$ being
the number of states of $Q_{i}$, $i=1,2$.

\end{Th}

\noindent{\bf Proof.} In this algorithm, we need a few symbols.
First, $\Sigma^{(k)}_{<}$ is defined as $\Sigma^{(k)}$ but it is
further required that the elements of $\Sigma^{(k)}$ are arranged
according to the  order $<$ from the least $\sigma_{1}^{k}$ to the
biggest $\sigma_{m}^{k}$, where $\sigma^{k}$ means the string
$\underbrace{\sigma\sigma\cdots\sigma}_{k}$. In addition, denote
$x\Sigma_{<}=\{ x\sigma_{1}, x\sigma_{2},\ldots,x\sigma_{m}\}$ for
$x\in\Sigma^{*}$.

We outline the process for searching for ${\cal B}(w)$, for all
$w\in\Sigma^{(k-1)}$. Initially, we let
$\langle\eta|\overline\nu(w)\in {\cal B}(w)$ for $w\in
\Sigma^{(k-1)}$, and let {\it queue} be $\Sigma^{(k)}_{<}$. Then
take an element $x$ from $queue$ and  determine $x=yw_{i}$ for
some $y\in\Sigma^{*}$ and $w_{i}\in\Sigma^{(k-1)}$, and further
check whether or not $\langle\eta|\overline\nu(x)\in span {\cal
B}(w_i)$. If $\langle\eta|\overline\nu(x)\not\in span {\cal
B}(w_i)$, then add $\langle\eta|\overline\nu(x)$ to ${\cal
B}(w_i)$ and  add the elements of $ x \Sigma_{<}$ to {\it queue}
from small element to large one sequentially. By repeating the
above process, we take another element $x'$ from {\it queue} and
then determine $x'=y'w_{j}$ for some $y'\in\Sigma^{*}$ and
$w_{j}\in\Sigma^{(k-1)}$, and further check whether or not
$\langle\eta|\overline\nu(x')\in span {\cal B}(w_j)$. Continue
this process, until  {\it queue} is empty. Since the number of the
elements in $\bigcup_{w\in\Sigma^{(k-1)}}{\cal B}(w)$ is at most
$n^{2}m^{k-1}$, the number of the elements in {\it queue} is at
most $n^{2}m^{k}$. In addition, from the proof of Lemma 1, we know
that $\mathbb{G}(l_{0}, w)=\mathbb{G}(l_{0}+i, w)$ for any $i\geq
0$. Therefore, if $x$ belongs to the $queue$, then $k-1\leq
|x|\leq l_{0}+k-1\leq (n^{2}-1)m^{k-1}+k$. Therefore, the above
process will always end.

We describe the process by Figure 1 and then we will analyze its
time complexity.

We prove that ${\cal B}(w)$ found out in the algorithm is exactly
a base of $\mathbb{G}(l_{0}, w)$. By the algorithm, the vectors in
${\cal B}(w)$ are linearly independent. Next, we only need to
prove that, for every $w\in \Sigma^{(k-1)}$, all vectors in
$\mathbb{G}(l_0,w)$ can be linearly expressed by the vectors in
${\cal B}(w)$. It suffices to prove that, for  any
$x\in\Sigma^{*}$ with $x=x_0 w$, $\langle\eta|\overline{\nu}(x)$
can be linearly expressed by the vectors in ${\cal B}(w)$. We
proceed by induction on the partial order $\leq$ on $\Sigma^{*}$.

(1) \emph{Basis.}

When $x_{0}=\epsilon$, we know that
$\langle\eta|\overline{\nu}(w)\in {\cal B}(w)$ by the beginning of
the algorithm. Thereby, $\langle\eta|\overline{\nu}(x)$ can be
linearly expressed by the vectors in ${\cal B}(w)$.

\begin{center}
 Figure 1.
 Algorithm (I) for searching for ${\cal B}(w)$ for all $w\in\Sigma^{(k-1)}$.
 \fbox{\parbox{\textwidth}{
 {\small\vskip 1mm
\begin{quote} {\bf Input:} ${\cal
A}_1=(Q_{1},Q_{acc}^{(1)},|\psi_{0}^{(1)}\rangle, \Sigma,\mu_{1})$
and   ${\cal A}_2=(Q_{2},Q_{acc}^{(2)},|\psi_{0}^{(2)}\rangle,
\Sigma,\mu_{2})$
\begin{quote}
Set ${\cal B}(w)=\{\langle\eta|\overline\nu(w)\}$, for $w\in\Sigma^{(k-1)}$;              \\
 $queue\leftarrow \Sigma^{(k)}_{<}$;\\
 {\bf while} $queue$ is not empty {\bf do}\\
 {\bf begin} take an element $x$ from $queue$;
 \begin{quote}
{\bf if} $x=yw$ for some $y\in\Sigma^{*}$ and
$\langle\eta|\overline\nu(x) \notin span {\cal B}(w)$ {\bf
then} \\
{\bf begin}  add $\langle\eta|\overline\nu(x)$ to ${\cal B}(w)$;
\begin{quote}
 add  all elements of $x\Sigma_{<} $
to $queue$ from the least $x\sigma_{1}$ to the biggest
$x\sigma_{m}$ sequentially;
\end{quote}
{\bf end};
 \end{quote}
 {\bf end}
\end{quote}

\end{quote}
}
 }
 }
\end{center}

(2) \emph{Induction.}

Assume that, for for some $y > \epsilon$, when $x_{0}<y$,
$\langle\eta|\overline{\nu}(x_{0}w)$ can be linearly expressed by
the vectors in ${\cal B}(w)$.  Now, we prove that
$\langle\eta|\overline{\nu}(yw)$ can be linearly expressed by the
vectors in ${\cal B}(w)$ as well.

(i) If $x=yw$ appears in $queue$, then we can clearly know that
$\langle\eta|\overline{\nu}(x)$ can be linearly expressed by the
vectors in ${\cal B}(w)$ by  the algorithm.

(ii) If $x=yw$ does not appear in $queue$, then we know that there
exist  $w'\in \Sigma^{(k-1)}$ and $x_1, x_2\in\Sigma^*$ with
$|x_{2}|\geq 1$ such that
\begin{equation}
 x=yw=x_1 w' x_2,
\end{equation} and \begin{equation}
\langle\eta|\overline{\nu}(x_1
w')=\sum_{\gamma\in\Gamma}p_\gamma\langle\eta|\overline{\nu}(y_\gamma
w'),
\end{equation}
for  some set of indices $\Gamma$,  where $p_\gamma\in{\mathbb
C}$, $y_\gamma\in \Sigma^*$, and $y_\gamma<x_1$ for all $\gamma
\in\Gamma$. Therefore,
\begin{equation}
\langle\eta|\overline{\nu}(x)=\sum_{\gamma\in\Gamma}p_\gamma\langle\eta|\overline{\nu}(y_\gamma
w' x_2) \label{algorithm}
\end{equation}
and $y_\gamma w' x_2<x_{1}w' x_2=x=yw$. Let $y_\gamma w'
x_2=y_{\gamma}'w$. Then $y_\gamma'<y$. Therefore, by the inductive
assumption, we know that $\langle\eta|\overline{\nu}(y_\gamma' w)$
can be linearly expressed by the vectors in ${\cal B}(w)$.
Consequently, by Eq. (\ref{algorithm}) we obtain that
$\langle\eta|\overline{\nu}(x)$ can be linearly expressed by the
vectors in ${\cal B}(w)$.

(3) \emph{Conclusion.}

For any $\langle\eta|\overline{\nu}(x)\in\mathbb{G}(l_0,w)$,
$\langle\eta|\overline{\nu}(x)$ can be linearly expressed by the
vectors in ${\cal B}(w)$. Hence, ${\cal B}(w)$ is a base of
$\mathbb{G}(l_0,w)$, for every $w\in\Sigma^{(k-1)}$.

\paragraph{Complexity of the algorithm.}
First we assume that all the inputs consist of complex numbers
whose real and imaginary parts are rational numbers and that each
arithmetic operation on rational numbers can be done in constant
time. Note that with time $O(|x|n^{4})$ we compute
$\langle\eta|\overline\nu(x)$. Then we need to recall that  to
verify whether a set of $n$-dimensional vectors is linearly
independent needs time $O(n^{3})$ \cite{FF63}. The time complexity
to check whether or not $\langle\eta|\overline\nu(x) \in span
{\cal B}(w)$ is $O(n^{6})$.

Because the basis ${\cal B}(w)$ has at most $ n^2$ elements,
$\bigcup_{w\in\Sigma^{(k-1)}}{\cal B}(w)$ has at most $n^2
m^{k-1}$ elements. Every element produces at most $m$ valid
child nodes. Therefore, we visit at most $O(n^2 m^{k})$ nodes
in {\it queue}. At every visited node $x$ the algorithm may do
three things: (i) calculating $\langle\eta|\overline{\nu}(x)$,
which needs time $O(|x|n^{4})$; (ii) finding the string $w$
composed of the last $k-1$ letters of $x$, which needs time
$O(k)$; (iii) verifying whether or not the $n^2$-dimensional
vector $\langle\eta|\overline{\nu}(x)$ is linearly independent of
the set ${\cal B}(w)$, which needs time $O(n^6)$ according to the
result in \cite{FF63}. In addition, from the proof of Lemma 1, we
know that $\mathbb{G}(l_{0}, w)=\mathbb{G}(l_{0}+i, w)$ for any
$i\geq 0$ and any $w\in\Sigma^{(k-1)}$. Therefore, if $x$ belongs
to the $queue$, then $k\leq |x|\leq l_{0}+k-1\leq
(n^{2}-1)m^{k-1}+k$.

Hence, the worst time complexity is
$O(n^2m^k(n^6m^{k-1}+kn^{4}+k+n^6))$, that is,
$O(m^{2k-1}n^{8}+km^kn^{6})$. \qed\\

\subsection{A polynomial-time algorithm for determining the equivalence between multi-letter
QFAs}

By virtue of Algorithm (I), we can determine the equivalence
between any two multi-letter QFAs in polynomial time.

\begin{Th}

Let  $\Sigma$,  $k_{1}$-letter QFA ${\cal A}_1$, and $k_{2}$-letter
QFA ${\cal A}_2$ be the same as Theorem 7. Then there exists a
polynomial-time $O(m^{2k-1}n^{8}+km^kn^{6})$ algorithm to determine
whether or not ${\cal A}_1$ and ${\cal A}_2$ are equivalent.

\end{Th}

\noindent{\bf Proof.} Denote ${\cal
B}_{0}=\{\langle\eta|\overline\nu(x): x\in
\bigcup_{l=0}^{k-2}\Sigma^{(l)} \}$. By using Algorithm (I), in time
$O(m^{2k-1}n^{8}+km^kn^{6})$ we can find
$\bigcup_{w\in\Sigma^{(k-1)}}{\cal B}(w)$. Since ${\cal B}(w)$ is a
base of $\mathbb{G}(l_{0}, w)$, we know that every vector in
$\{\langle\eta|\overline\nu(x): x\in\Sigma^{*}, |x|\geq 1 \}$ can be
linearly expressed by a finite number of vectors in
$\bigcup_{w\in\Sigma^{(k-1)}} {\cal B}(w) \bigcup  {\cal B}_{0}$.
Therefore, we have the following algorithm.

\begin{center}
 Figure 2.
 Algorithm (II) for determining the equivalence between multi-letter QFAs. \vskip 1mm
 \fbox{\parbox{\textwidth}{
 {\small\vskip 1mm
{\bf Input:} ${\cal
A}_1=(Q_{1},Q_{acc}^{(1)},|\psi_{0}^{(1)}\rangle,
\Sigma,\mu_{1},k_{1})$ and ${\cal
A}_2=(Q_{2},Q_{acc}^{(2)},|\psi_{0}^{(2)}\rangle,
\Sigma,\mu_{2},k_{2})$.\\
{\bf Step 1:}  \begin{quote} By means of Algorithm (I), find out
${\cal B}(w)$ for all $w\in\Sigma^{(k-1)}$;\\ compute ${\cal
B}_{0}$;
\end{quote}
{\bf Step 2:}
 \begin{quote} {\bf If} $\forall \langle\psi|\in
\bigcup_{w\in\Sigma^{(k-1)}}{\cal B}(w) \bigcup {\cal B}_{0}$,
$\langle\psi|P_{acc}\rangle= 0$ {\bf then} return
(${\cal A}_1$ and ${\cal A}_2$ are  equivalent)\\
{\bf else} return the $x$ for which
$\langle\eta|\overline\nu(x)|P_{acc}\rangle\neq 0$;
\end{quote}

}}}
\end{center}

\paragraph{Complexity of the algorithm.}  We
compute ${\cal B}_{0}$ in time $O(m^{k-2}n^{4})$. Therefore, by
Algorithm (I), in Step 1, we need time $O(m^{2k-1}n^{8}+km^kn^{6})$.

Provided $\langle\eta|\overline\nu(x)$ has been computed, in time
$O(n^{2})$ we can decide whether
$\langle\eta|\overline\nu(x)P_{acc}\neq 0$. Since the number of
elements in ${\cal B}_{0}\cup \bigcup_{w\in\Sigma^{(k-1)}} {\cal
B}(w)$ is at most $O(m^{k-1}n^{2})$, the worst time complexity in
Step 2 is  $O(m^{k-1}n^{4})$.

In summary,  the worst time complexity in Algorithm (II) is
$O(m^{2k-1}n^{8}+km^kn^{6})$. \qed\\

\section{Concluding remarks and some open problems}

Because the method of decidability of equivalence for multi-letter QFAs with single input alphabet in \cite{QY08} is not applied to the case of multi-input alphabet directly,
in this paper, we have solved the decidability of equivalence of
multi-letter QFAs.  More exactly, we have proved that any two
automata, a $k_{1}$-letter QFAs ${\cal A}_1$ and a $k_{2}$-letter
${\cal A}_2$ over the input alphabet
$\Sigma=\{\sigma_{1},\sigma_{2},\ldots,\sigma_{m}\}$, are
equivalent if and only if they are
$(n^2m^{k-1}-m^{k-1}+k)$-equivalent, where $n=n_{1}+n_{2}$,
$n_{1}$ and $n_{2}$ are the numbers of states of ${\cal A}_{1}$
and ${\cal A}_{2}$, respectively, and $k=\max(k_{1},k_{2})$.  By
this result, we have obtained the decidability of equivalence of
multi-letter QFAs over the same single input alphabet
$\Sigma=\{\sigma\}$ and the decidability of equivalence of
MO-1QFAs.

However, if we determine the equivalence of multi-letter QFAs by
checking all strings $x$ with $|x|\leq n^2m^{k-1}-m^{k-1}+k$, then
the time complexity is exponential. Therefore, we have designed a
polynomial-time $O(m^{2k-1}n^{8}+km^kn^{6})$ algorithm for
determining the equivalence of any two multi-letter QFAs.

It is worth pointing out that although we have given an upper
bound on  the length of strings to be verified for determining
whether two multi-letter QFAs are equivalent, the optimality of
the upper bound has not been discussed, and this is worthy of
further consideration. An open issue concerns the state complexity
of multi-letter QFAs compared with the MO-1QFAs for accepting some
languages (for example, unary regular languages \cite{RS97,Yu98}).
Also, recalling the relation between MM-1QFAs and MO-1QFAs, the
power of measure-many multi-letter QFAs is worth being clarified.
Whether or not measure-many multi-letter QFAs can recognize
non-regular languages may also be considered in the future. Finally, with the equivalence result in the paper, we may consider the minimization of states regarding multi-letter QFAs.

\par

\end{document}